\documentclass[12pt,letterpaper]{article}

\usepackage{natbib,hyperref}

\usepackage[ left=1in, top=1in, right=1in, bottom=1in]{geometry}
\usepackage{graphicx,bm,colonequals,amsmath,amssymb,url,xcolor,bbm}
\usepackage{array,tabularx,multirow}
\usepackage{enumitem,MnSymbol}
\usepackage[font={footnotesize}]{caption,subcaption}
\usepackage[utf8]{inputenc}
\usepackage{soul} 
\usepackage{placeins} 
\usepackage[normalem]{ulem}

\graphicspath{{plots/}}

\usepackage{mathtools}
\mathtoolsset{showonlyrefs} 

\bibpunct[, ]{(}{)}{;}{a}{,}{,}
   
\usepackage{amsthm}
\newtheoremstyle{propstyle} 
    {2mm}                    
    {1mm}                    
    {\itshape}                   
    {}                           
    {\scshape}                   
    {.}                          
    {.5em}                       
    {}  
\theoremstyle{propstyle}

\theoremstyle{propstyle}

\theoremstyle{propstyle}

\theoremstyle{propstyle}

\theoremstyle{propstyle}

\usepackage{array,framed}
\newcounter{algorithm}

\makeatletter
\renewcommand{\paragraph}{%
  \@startsection{paragraph}{4}%
  {\z@}{2ex \@plus 1ex \@minus .2ex}{-1em}%
  {\normalfont\normalsize\bfseries}%
}
\makeatother

\usepackage{etoolbox}
\makeatletter
\preto{\@verbatim}{\topsep=0pt \partopsep=0pt }
\makeatother

\DeclareMathAlphabet\mathbfcal{OMS}{cmsy}{b}{n}

\newcommand{\bx}{\mathbf{x}}
\newcommand{\by}{\mathbf{y}}

\newcommand{\bg}{\mathbf{g}}

\newcommand{\bK}{\mathbf{K}}

\newcommand{\bM}{\mathbf{M}}

\newcommand{\bflambda}{\bm{\lambda}}
\newcommand{\bfmu}{\bm{\mu}}

\newcommand{\bftheta}{\bm{\theta}}

\newcommand{\bfbeta}{\bm{\beta}}

\newcommand{\bfpsi}{\bm{\psi}}

\newcommand{\GP}{\mathcal{GP}}

\newcommand{\order}{\mathcal{O}}
\newcommand{\normal}{\mathcal{N}}

\newcommand{\dens}{p}
\newcommand{\adens}{\widehat p}

\newcommand{\domain}{\mathcal{X}}

\title{Scaled Vecchia approximation for fast computer-model emulation}

\author{Matthias Katzfuss\thanks{Department of Statistics, Texas A\&M University. Corresponding author: \texttt{katzfuss@gmail.com}} 
\and Joseph Guinness\thanks{Department of Statistics and Data Science, Cornell University} 
\and Earl Lawrence\thanks{Los Alamos National Laboratory}}

\date{}

\begin{document}

\maketitle

\begin{abstract}
Many scientific phenomena are studied using computer experiments consisting of multiple runs of a computer model while varying the input settings. Gaussian processes (GPs) are a popular tool for the analysis of computer experiments, enabling interpolation between input settings, but direct GP inference is computationally infeasible for large datasets. We adapt and extend a powerful class of GP methods from spatial statistics to enable the scalable analysis and emulation of large computer experiments. Specifically, we apply Vecchia's ordered conditional approximation in a transformed input space, with each input scaled according to how strongly it relates to the computer-model response. The scaling is learned from the data, by estimating parameters in the GP covariance function using Fisher scoring. Our methods are highly scalable, enabling estimation, joint prediction and simulation in near-linear time in the number of model runs. In several numerical examples, our approach substantially outperformed existing methods.
\end{abstract}


\section{Introduction \label{sec:intro}}


At the cutting edge of science, computationally intensive simulations are used to make predictions of complex phenomena, such as the distribution of matter in the Universe \citep{lawrence2017mira}, the behavior of materials under high pressure \citep{walters2018bayesian}, or the composition of rocks on Mars \citep{bhat2020estimating}. 
These simulations are simply too slow for use in data analysis \citep{higdon2004combining} or real-time applications \citep{mehta2014modeling}, so the statistics discipline known as computer experiments has grown to address this computational challenge. The key ingredient in much of this work is an emulator, a statistical approximation to the computer simulation. Emulators can predict the output of a simulation many orders of magnitude faster than the simulation itself, at the cost of additional error. 
Emulation is achieved by building a regression model from the inputs to the outputs.


Gaussian processes (GPs) are popular emulators and have emerged as indispensable tools for design, analysis, and calibration of computer experiments \citep[e.g.,][]{Sacks1989,Kennedy2001}.
GPs are accurate, flexible, interpretable, and probabilistic, thus providing natural quantification of uncertainty. 
For the analysis of $n$ computer-model runs, GP inference typically requires working with a dense $n \times n$ covariance matrix. Thus, direct GP inference is infeasible for many present and future computer experiments, as new supercomputers enable increasingly large numbers of increasingly detailed simulations to be carried out.
Scalability improvements for computer-experiment methods are vital to handle the expected increase in simulation output. 


Many approaches have been proposed to enable scalable GP inference. \citet{Heaton2017} review and compare approaches from spatial statistics, and \citet{Liu2018} review approaches in machine learning. In the context of large computer experiments, scalable GP approaches include compactly supported covariances \citep{Kaufman2011}; sparse grid-based GPs \citep{plumlee2014fast}; and the local approximate GP (laGP) of \citet{Gramacy2015}, which makes independent predictions at different input values using nearby observations in the input space.
In spatial statistics, the Vecchia approximation \citep{Vecchia1988} and its extensions \citep[e.g.,][]{Stein2004,Datta2016,Guinness2016a,Katzfuss2017a,Katzfuss2018} are very popular GP approximations. Similar to the laGP, the Vecchia approximation considers nearest neighbors, but it does so from an ordered conditional perspective; as a result, Vecchia approximations imply a valid joint distribution for the data, resulting in  straightforward global likelihood-based parameter inference and joint predictions at a set of input values, which allows proper uncertainty quantification in down-stream applications.


Here, we adapt and extend the powerful class of Vecchia GP approximations from spatial statistics to enable the scalable analysis and emulation of large computer experiments. Specifically, we apply Vecchia's ordered conditional approximation in a transformed input space, for which each input is scaled according to how strongly it is related to the computer-model response. The scaling of the input space is learned from the data, by estimating parameters in the GP covariance function using Fisher scoring \citep{Guinness2019}. Our scaled Vecchia methods are highly scalable, enabling ordering, neighbor-search, estimation, joint prediction and simulation in near-linear time in the number of model runs. Thus, our methods can handle large numbers of model runs, joint sampling of paths through the input space, and relatively high input dimensions, assuming that only some of the inputs have a strong effect on the output, while others are less important.
Recently, \citet{Shi2017} also applied a Vecchia-type approximation in a GP-emulation setting, but their focus was on infinitely smooth covariances (i.e., squared exponential) and change-of-support problems.
\citet{Datta2016a} proposed a Bayesian Vecchia-type approximation for spatio-temporal processes, in which the neighbors (but not the ordering) were selected based on the spatio-temporal correlations; for a specific choice of covariance function, their correlation-based neighbor-selection procedure can be viewed as neighbor-selection based on distances between scaled inputs, similar as in our approach here.


Relative to recent work on Vecchia approximations of spatial processes by the authors \citep{Guinness2016a,Katzfuss2017a,Katzfuss2018,Guinness2019}, the present paper makes several contributions that are crucial to addressing challenges with computer-model emulation. We extend the Vecchia approximation to non-spatial inputs and to anisotropic covariance functions. The Vecchia ordering and neighbor sets are determined based on scaled inputs; this greatly improves the accuracy in the high-dimensional input spaces common in computer experiments (as opposed to the usual two-dimensional space in spatial statistics).
When estimating the (unknown) scaling parameters in an iterative fashion, the scaling of the inputs changes along with the parameter estimates over the course of the Fisher-scoring iterations; hence, as we refine the estimates of the parameters, we refine our Vecchia approximation of the implied anisotropic covariance.
As a stationary GP may be less appropriate for modeling some computer-model surfaces than many geospatial fields, model misspecification and resulting underestimation of prediction uncertainty may be an issue; 
to address this, we developed a simple, computationally cheap, and effective variance-correction approach, resulting in well-calibrated and sharp predictive distributions. 

The remainder of this document is organized as follows. In Section \ref{sec:review}, we describe GP emulation of computer models, and we review existing Vecchia approximations from spatial statistics. In Section \ref{sec:methodology}, we introduce our new scaled Vecchia methods for fast emulation of large computer experiments. In Section \ref{sec:comparison}, we provide numerical comparisons to existing approaches. Section \ref{sec:conclusions} concludes and discusses future work. \texttt{R} code to run our method and reproduce all results is available at \url{https://github.com/katzfuss-group/scaledVecchia}.

\section{Review\label{sec:review}}

\subsection{Computer-model emulation using Gaussian processes\label{sec:gp}}

Let $y(\bx)$ be the response of a computer model at a $d$-dimensional input vector $\bx$ on the input domain $\domain$. In Gaussian-process emulation, $y(\cdot)\sim \GP(\mu,K)$ is assumed to be a Gaussian process (GP) with mean function $\mu: \domain \rightarrow \mathbb{R}$ and a positive-definite covariance or kernel function $K: \domain \times \domain \rightarrow \mathbb{R}$. Then, the vector $\by = \big(y(\bx_1),\ldots,y(\bx_n)\big)^\top$ of responses at $n$ input values $\{\bx_1,\ldots,\bx_n\}$ follows an $n$-variate Gaussian distribution with covariance matrix
$
\bK = \big(K(\bx_i,\bx_j)\big)_{i,j=1,\ldots,n},
$ 
whose $(i,j)$th entry describes the covariance between the responses of simulations $i$ and $j$ as a function of the corresponding input values $\bx_i$ and $\bx_j$.

For simplicity, we henceforth make some additional assumptions, although most of our methodology is also applicable in more general situations. Specifically, we assume that the mean function $\mu(\bx) = \bfpsi(\bx)^\top \bfbeta$ is linear in a number of covariate parameters, $\bfbeta$; typical assumptions are $\bfpsi(\bx) = 1$ or $\bfpsi(\bx) = (1,\bx^\top)^\top$.

We also assume an anisotropic covariance function with a separate range parameter $\lambda_l$ for each input dimension $l$, also referred to as automatic relevance determination: 
$K(\bx_i,\bx_j) = \tilde K(q(\bx_i,\bx_j))$,
where
\begin{equation}
    \label{eq:scaleddist}
\textstyle q(\bx_i,\bx_j) 
= \big(\, \sum_{l=1}^d (\frac{x_{i,l}-x_{j,l}}{\lambda_l})^2 \,\big)^{1/2},
\end{equation}
and $\tilde K$ can be any covariance function that is valid (i.e., strictly positive definite) in $\mathbb{R}^d$, such as the squared exponential or Mat\'ern covariance function.

While GPs are indispensable tools for computer-model emulation due to their flexibility and natural uncertainty quantification, direct GP inference requires an $\order(n^3)$ factorization of the covariance matrix, which is not feasible for large computer experiments. Thus, we propose an approximation that reduces computational complexity and hence improves scalability.

\subsection{Vecchia approximations in spatial statistics\label{sec:spatial}}

Vecchia's approximation \citep{Vecchia1988} is a powerful GP approximation that is popular in spatial statistics. Motivated by the exact decomposition of the joint density $\dens(\by) = \prod_{i=1}^n \dens(y_i|\by_{1:i-1})$ as a product of univariate conditional densities, \citet{Vecchia1988} proposed the approximation
\begin{equation}
    \label{eq:vecchia}
\textstyle \adens(\by) = \prod_{i=1}^n \dens(y_i|\by_{c(i)}),
\end{equation}
where $c(i) \subset \{1,\ldots,i-1\}$ is a conditioning index set of size $|c(i)| = \min(m,i-1)$ for all $i=2,\ldots,n$ (and $c(1) = \emptyset$). Even with relatively small conditioning-set size $m\ll n$, the approximation \eqref{eq:vecchia} with appropriate choice of the $c(i)$ can often be very accurate due to the screening effect \citep[e.g.,][]{Stein2011b}. The $\dens(y_i|\by_{c(i)})$ in \eqref{eq:vecchia} are all Gaussian distributions that can be computed in parallel using standard formulas, each using $\order(m^3)$ operations based on $\order(m)$ data.

The Vecchia approximation has many useful properties. For example, the implied joint distribution $\adens(\by) = \normal_n(\bfmu,\widehat\bK)$ is also multivariate Gaussian, and the Cholesky factor of $\widehat\bK^{-1}$ is highly sparse with fewer than $nm$ off-diagonal nonzero entries \citep[e.g.,][]{Datta2016,Katzfuss2017a}. 
Further, under the sparsity constraint implied by the choice of the $c(i)$, the Vecchia approximation results in the optimal inverse Cholesky factor $\widehat\bK^{-1/2}$, as measured by the Kullback-Leibler (KL) divergence, $KL(\dens(\by)\| \adens(\by))$ \citep{Schafer2020}. 
Enlarging the conditioning sets $c(i)$ never increases the KL divergence \citep{Guinness2016a}; for $m=n-1$, the approximation is exact, $\adens(\by) = \dens(\by)$.
In contrast to local GP approximations, the Vecchia approximation to the underlying model is global; thus, for example, model parameters can be estimated (see Section \ref{sec:estimation}) from a subsample of the data, and then the estimated parameters can be used to make predictions (Section \ref{sec:pred}) using all of the data.

The approximation accuracy of the Vecchia approach depends on the choice of ordering of the variables $y_1,\ldots,y_n$ and on the choice of the conditioning sets $c(i)$. A general Vecchia framework \citep{Katzfuss2017a} obtained by varying these choices unifies many popular GP approximations \citep[e.g.,][]{Quinonero-Candela2005, Snelson2007, Banerjee2008,Katzfuss2015,Katzfuss2017b}.
In practice, high accuracy can be achieved using a maximum-minimum distance (maximin) ordering and nearest-neighbor (NN) conditioning, which are illustrated in Figure \ref{fig:ordero}. Maximin ordering picks the first variable arbitrarily, and then chooses each subsequent variable in the ordering as the one that maximizes the minimum distance to previous variables in the ordering. For NN conditioning, each $c(i)$ then consists of the indices corresponding to the $m$ nearest previously ordered variables. For both ordering and conditioning, distance between two variables $y_i$ and $y_j$ is typically defined as the Euclidean distance $\|\bx_i - \bx_j\|$ between their corresponding inputs. In addition, we employ a grouping strategy \citep{Guinness2016a} that combines conditioning sets $c(i)$ and $c(j)$ when doing so is computationally advantageous.
When using maximin ordering and NN conditioning, recent results \citep{Schafer2020} imply that, for increasing $n$, a specific accuracy for certain isotropic Mat\'ern kernels can be guaranteed using conditioning sets of size $m=\order(\log^d(n))$, under regularity conditions and ignoring edge effects. The resulting near-linear time complexity is the best known complexity for problems of this type.

\section{Methodology\label{sec:methodology}}

Several issues arise when applying Vecchia approximations for spatial GPs to emulation of computer experiments (Section \ref{sec:gp}). While physical distance between spatial locations is usually meaningful, Euclidean distance between inputs to a computer experiments depends heavily on the arbitrary scaling of each input dimension. In addition, while spatial fields are typically two- or three-dimensional, computer experiments often consider $d \approx 10$ inputs; as the asymptotics discussed at the end of Section \ref{sec:spatial} imply that $m=\order(\log^d(n))$ is required to achieve a certain accuracy, a very large $m$ might be required for large $d$, resulting in a prohibitive computational cost (which scales cubically in $m$).

\subsection{Scaled Vecchia approximation for computer experiments}

\begin{figure}
\centering
	\begin{subfigure}{.28\textwidth}
	\centering
	\includegraphics[width =.9\linewidth]{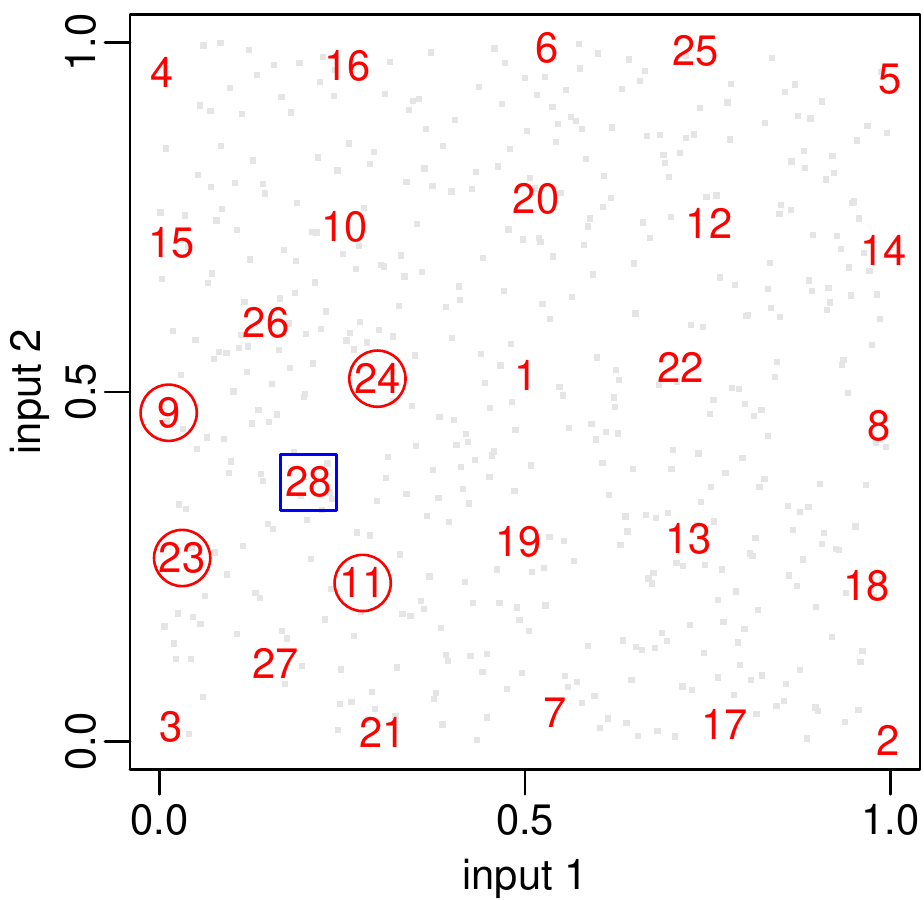}
	\caption{MN of $\bx_1,\ldots,\bx_n$, shown in $\domain$}
	\label{fig:ordero}
	\end{subfigure}%
	\begin{subfigure}{.7\textwidth}
	\centering
	\includegraphics[width =.98\linewidth]{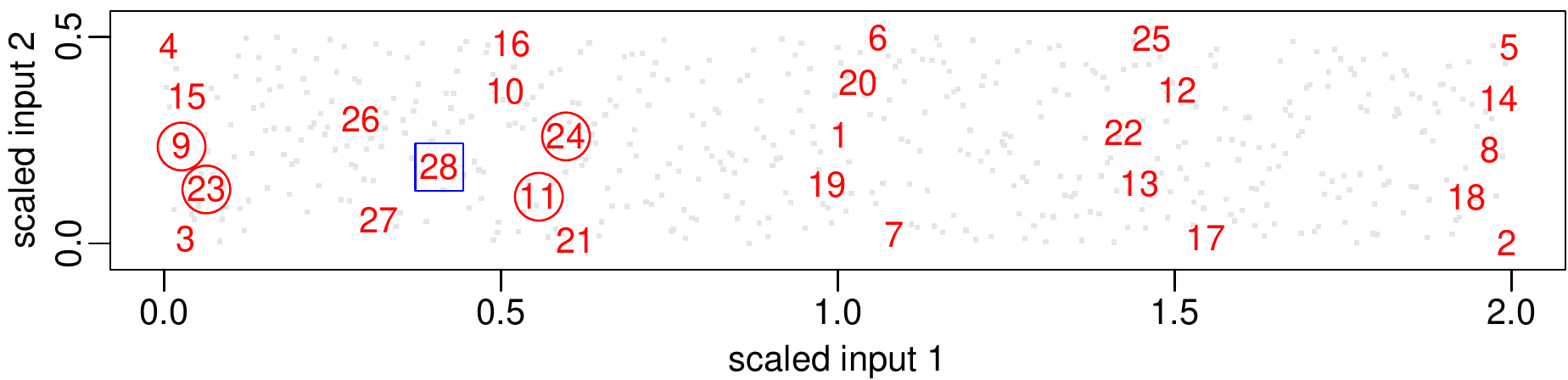}
	\caption{MN of $\bx_1,\ldots,\bx_n$, shown in $\tilde\domain$}
	\label{fig:orders}
	\end{subfigure} \\
	\vspace{3mm}
	\begin{subfigure}{.7\textwidth}
	\centering
	\includegraphics[width =.9\linewidth]{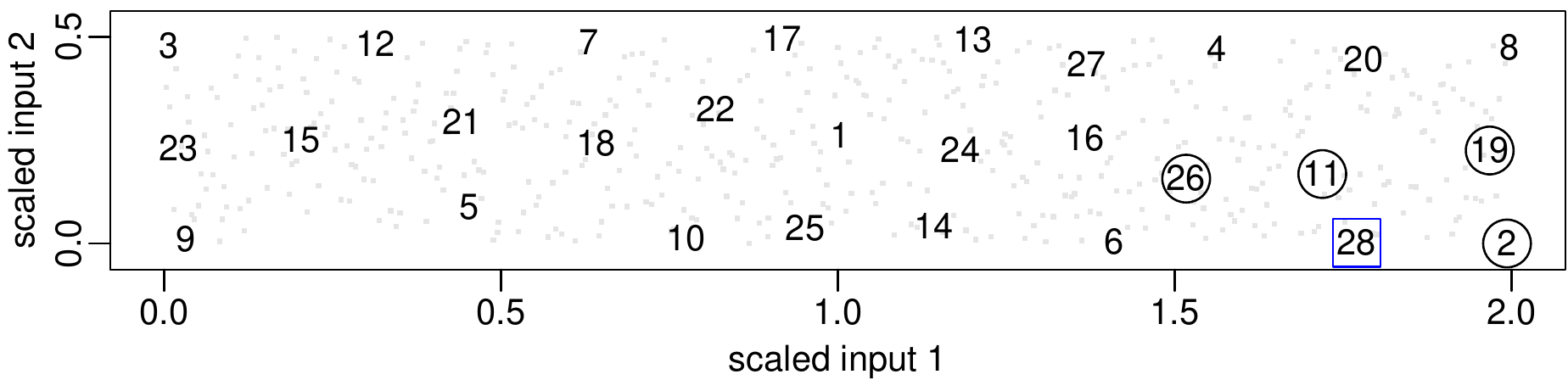}
	\caption{MN of $\tilde\bx_1,\ldots,\tilde\bx_n$, shown in $\tilde\domain$}
	\label{fig:orderss}
	\end{subfigure}%
  \caption{Maximin ordering and nearest-neighbor conditioning (MN) for $n=500$ inputs (small grey points) generated using Latin hypercube sampling on $\domain = [0,1]^2$ in $d=2$ dimensions, assuming an anisotropic covariance \eqref{eq:scaleddist} with range parameters $\bflambda=(1/2,2)$. 
  MN is carried out on the original inputs $\bx_1,\ldots,\bx_n$ (top row, red) or the scaled inputs $\tilde\bx_1,\ldots,\tilde\bx_n$ (bottom row, black).
  The first $i=28$ ordered inputs are numbered, with emphasis on the $i$th input (${\color{blue}\Box}$) and its nearest $m=4$ previously ordered neighbors with indices $c(i)$ (${\color{black}\bigcircle}$). 
  (a) MN of original inputs viewed on original space $\domain= [0,1] \times [0,1]$: First $i$ inputs are spread out over input space, $c(i) = (9,11,23,24)$ are nearby. (b) Same MN on scaled space $\tilde\domain = [0,2] \times [0,1/2]$: First $i$ inputs are irregularly spaced, $c(i)$ missed nearby $26$ and $27$. (c) MN of scaled inputs on scaled space: First $i$ inputs are spread out over input space, $c(i) = (2,11,19,26)$ are nearby, as desired.}
\label{fig:orderillus}
\end{figure}

We propose a scaled Vecchia approximation that exploits that the input variables can vary widely in the magnitude of their effect on the response; this is sometimes referred to as factor sparsity. Specifically, for known range parameters $\bflambda=(\lambda_1,\ldots,\lambda_d)^\top$, the anisotropic distance in \eqref{eq:scaleddist} can be viewed as a Euclidean distance between scaled inputs,
\begin{equation}
    \label{eq:scaledinputs}
q(\bx_i,\bx_j) = \| \tilde\bx_i - \tilde\bx_j \|,
\end{equation}
where $\tilde\bx = (x_1/\lambda_1,\ldots,x_d/\lambda_d)$ are the scaled inputs, and we call $1/\lambda_l$ the relevance of the $l$th input dimension or variable $x_l$ (assuming standardized input space $\domain=[0,1]^d$). Similar scaling ideas have been considered for other GP approximations of computer experiments \citep[e.g.,][]{Gramacy2016}.

Our scaled Vecchia approximation is defined as in \eqref{eq:vecchia}, except based on a maximin ordering and NN conditioning of the scaled inputs $\tilde\bx_1,\ldots,\tilde\bx_n$, assuming known $\bflambda$ for now; more precisely, we define the distance between variables $y_i$ and $y_j$ as the Euclidean distance $\| \tilde\bx_i - \tilde\bx_j\|$ between the corresponding scaled inputs (see Figure \ref{fig:orderss}), instead of $\|\bx_i-\bx_j\|$ in the standard Vecchia approximation. Note that this scaled Vecchia approximation can be viewed as a special case of correlation-based Vecchia (Kang and Katzfuss, in prep.). The ordering and conditioning can be computed in quasilinear time in $n$ \citep{Schafer2017,Schafer2020}.

The resulting scaled Vecchia approximation of the GP $y(\cdot)$ with anisotropic kernel $K$, can be viewed as a standard Vecchia approximation of a GP with isotropic kernel $\tilde K$ with scaled inputs $\tilde\bx$ in the scaled input space $\tilde\domain$. Importantly, Euclidean distance is only meaningful in $\tilde\domain$, not in $\domain$. Figure \ref{fig:orders} shows that maximin ordering of $\bx_1,\ldots,\bx_n$ can be highly irregular in $\tilde\domain$, and nearest-neighbor conditioning of $\bx_1,\ldots,\bx_n$ may miss important and nearby inputs in $\tilde\domain$. In contrast, scaled Vecchia (Figure \ref{fig:orderss}) is directly carried out in $\tilde\domain$; the resulting ordering is more regular, and the conditioning set $c(i)$ contains the nearest previously ordered neighbors, as desired to achieve good screening properties in the conditional distributions in \eqref{eq:vecchia}.

Many computer codes contain input variables $x_l$ that only weakly affect the response $y$; this can be captured in our model by a large $\lambda_l$, so that changes in $x_l$ only result in small changes in $\tilde x_l = x_l/\lambda_l$, and thus only minor changes in position in $\tilde\domain$. In the extreme case of $\lambda_l \rightarrow \infty$, the input variable $x_l$ is effectively eliminated from the model and the dimension $\tilde d$ of $\tilde \domain$ is smaller than the dimension $d$ of the original input space, and thus a smaller $m=\order(\log^{\tilde d}(n))$ is required to achieve a certain approximation accuracy. But even for large but finite range parameters, Figure \ref{fig:klcomp} shows that scaled Vecchia can achieve a certain accuracy with much smaller $m$ than standard Vecchia (see Section \ref{sec:matern} for more details).

\subsection{Estimation of parameters\label{sec:estimation}}

In practice, the parameters $\bfbeta$ in the mean function $\mu$ and parameters $\bftheta$ in the covariance function $K$ are unknown, including the range or scaling parameters $\bflambda$. We estimate these parameters by maximizing the logarithm of the Vecchia likelihood in \eqref{eq:vecchia}. This is challenging due to the potentially large number of parameters. Hence, we use a Fisher scoring algorithm \citep{Guinness2019}, which exploits first- and second-derivative information for fast convergence but preserves the $\order(nm^3)$ scaling of the Vecchia approximation. We briefly review this algorithm here, but refer to \citet{Guinness2019} for details.

Let $\ell(\bfbeta,\bftheta) = \log \adens_{\bfbeta,\bftheta}(\by)$, where $\adens_{\bfbeta,\bftheta}(\by)$ is the Vecchia approximation from \eqref{eq:vecchia} with $m=m_{\text{est}}$, except that we have now made explicit the dependence of the density on the parameters. Taking derivatives of the conditional densities in \eqref{eq:vecchia} is challenging; replacing them by joint distributions,
\begin{equation}
    \label{eq:fisherdens}
\textstyle \ell(\bfbeta,\bftheta) = \sum_{i=1}^n \big( \log \dens_{\bfbeta,\bftheta}(y_i,\by_{c(i)}) - \log \dens_{\bfbeta,\bftheta}(\by_{c(i)}) \big),
\end{equation}
enables the use of well-known formulas for the gradient and Fisher information of the Gaussian distributions $\dens_{\bfbeta,\bftheta}(y_i,\by_{c(i)})$ and $\dens_{\bfbeta,\bftheta}(\by_{c(i)})$. 
Because $\bfbeta$ appears linearly in the mean of the Gaussian distributions, we can profile out $\bfbeta$ using the closed-form expression for the generalized least squares estimator $\hat\bfbeta(\bftheta)$. Then, starting with an initial value $\bftheta^{(0)}$, Fisher scoring for $\bftheta$ proceeds for $k=0,1,2,\ldots$ as
\begin{equation}
    \label{eq:fisherupdate}
 \bftheta^{(k+1)} = \bftheta^{(k)} + (\bM^{(k)})^{-1} \bg^{(k)},
\end{equation}
where $\bg^{(k)} = \frac{\partial \ell(\hat\bfbeta(\bftheta),\bftheta)}{\partial \bftheta} |_{\bftheta= \bftheta^{(k)}}$ and $\bM^{(k)} = -\mathbb{E}\frac{\partial^2\ell(\hat\bfbeta(\bftheta),\bftheta)}{\partial\bftheta\partial\bftheta'}\big|_{\bftheta=\bftheta^{(k)}}$ can be computed based on \eqref{eq:fisherdens} as the sum of $n$ log-densities that are at most of dimensions $m+1$. The algorithm is terminated when the dot product between the step and the gradient $\bg^{(k)}$ is less than $10^{-4}$, obtaining the estimates $\hat\bftheta = \bftheta^{(k+1)}$ and $\hat\bfbeta = \hat\bfbeta(\hat\bftheta)$. 
In practice, a mild penalization term (e.g., to discourage variance parameters that are much larger than the sample variance of the training data) is added to \eqref{eq:fisherdens} to improve convergence. Also, when the Fisher-scoring step fails to increase the loglikelihood, the step is replaced by a line search along the gradient. This concludes the review of \citet{Guinness2019}.

In our scaled Vecchia approach, over the course of the Fisher-scoring iterations, the estimate of $\bftheta$ will change, and along with it, the scaled inputs $\tilde\bx = (x_1/\lambda_1,\ldots,x_d/\lambda_d)$, the resulting maximin ordering and NN conditioning, and the implied approximate density $\adens_{\bfbeta,\bftheta}(\by)$. 
For the purpose of computing the derivatives and the resulting parameter update in \eqref{eq:fisherupdate}, we ignore the dependence of the ordering and conditioning on $\bftheta$; instead, we update the ordering and conditioning separately given the current estimate $\bftheta^{(k)}$, but only at certain iterations, say $k=2,4,8,16,\ldots$, to avoid slowing the algorithm unnecessarily. Hence, as we refine the estimates of the parameters, we refine our Vecchia approximation of the implied covariance.
If it is of interest, we can carry out crude variable selection and eliminate inactive input dimensions by setting $\lambda_l = \infty$ if $\lambda_l^{(k)}$ is over a certain threshold (e.g., $10^3$). 

\begin{figure}
\centering
	\begin{subfigure}{.3\textwidth}
	\centering
	\includegraphics[width =.95\linewidth]{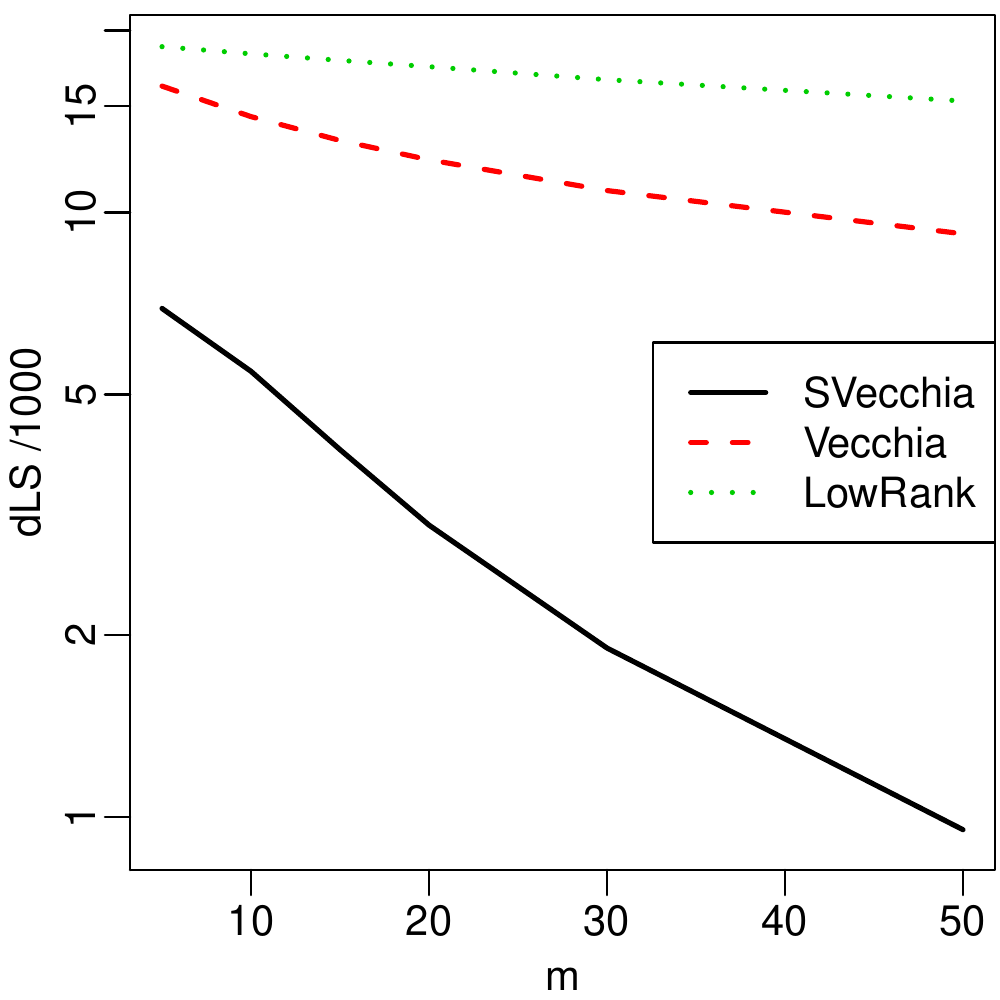}
	\caption{dLS (known parameters)}
	\label{fig:klcomp}
	\end{subfigure}%
	\begin{subfigure}{.3\textwidth}
	\centering
	\includegraphics[width =.95\linewidth]{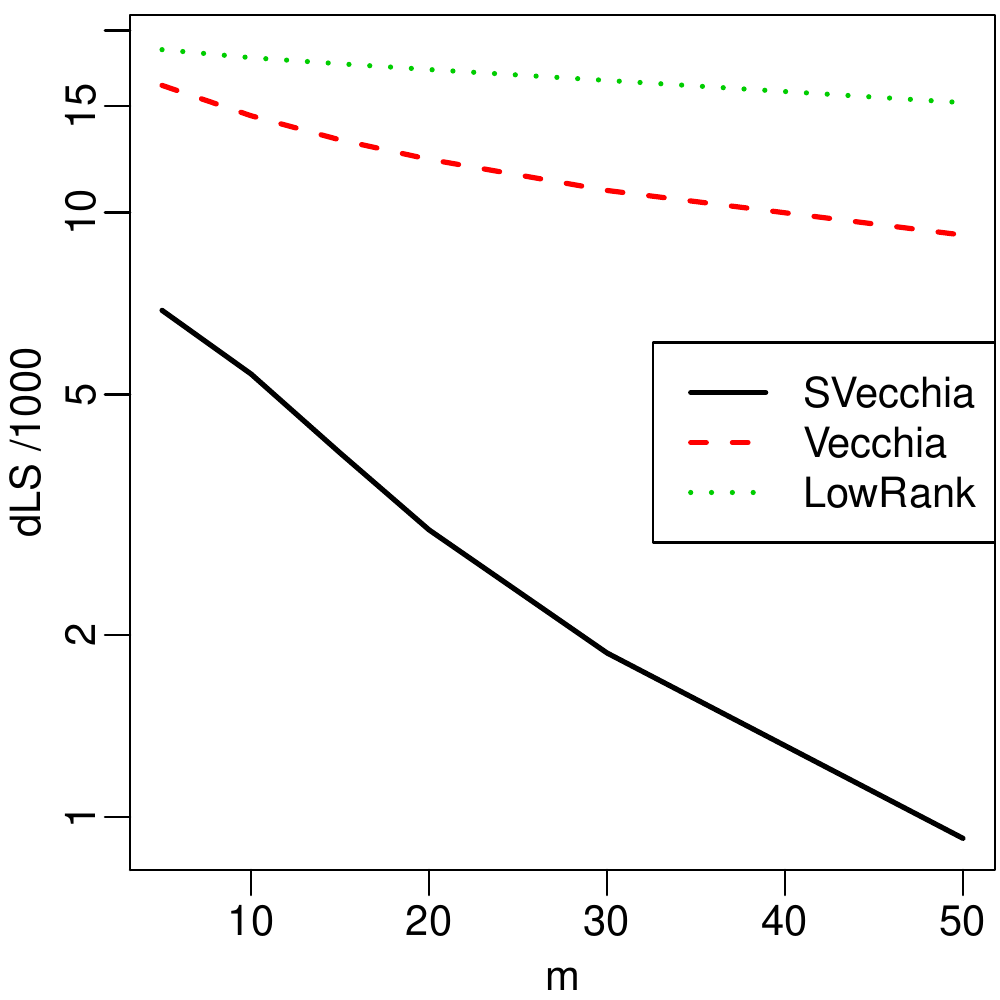}
	\caption{dLS (estimated parameters)}
	\label{fig:maternest}
	\end{subfigure}%
	\begin{subfigure}{.3\textwidth}
	\centering
	\includegraphics[width =.95\linewidth]{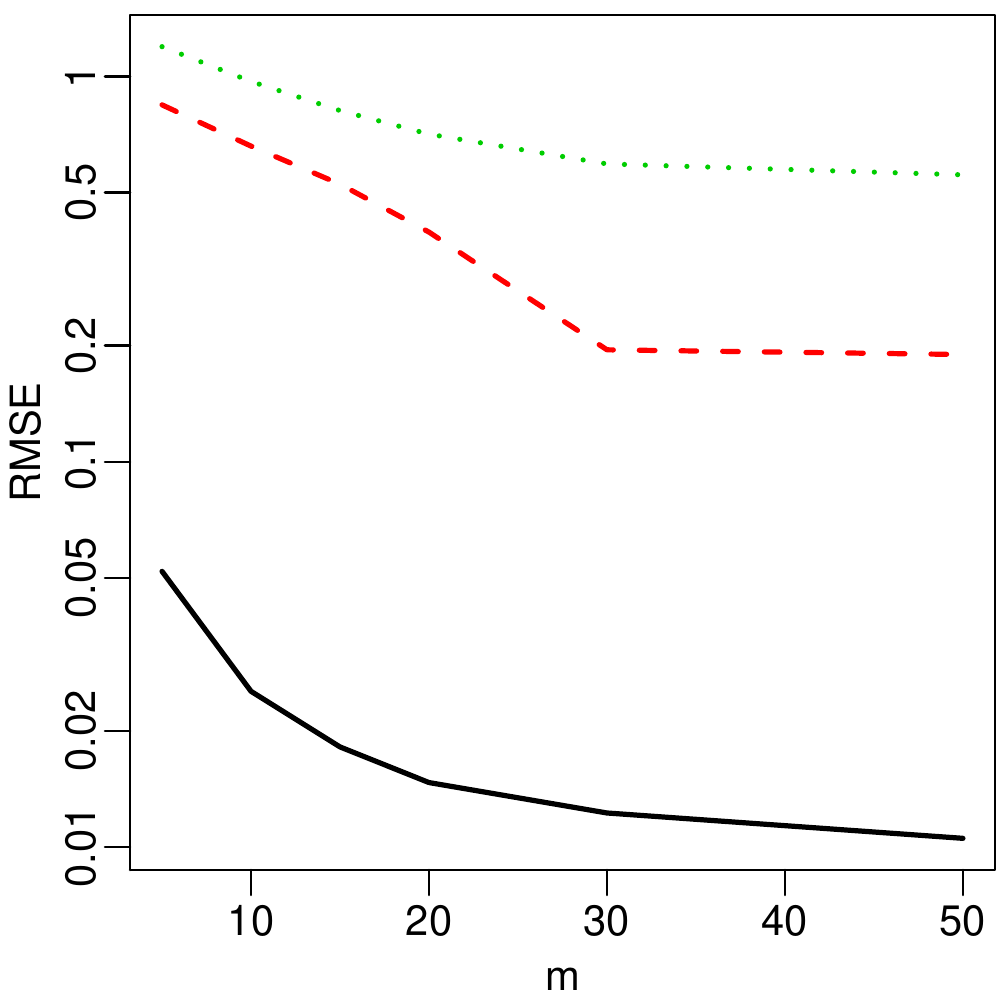}
	\caption{Prediction RMSE}
	\label{fig:maternpred}
	\end{subfigure}%
  \caption{For data simulated from a Mat\'ern GP in $d=10$ input dimensions, comparison of our proposed scaled Vecchia (SVecchia) approach to two existing GP approximations, in terms of average difference in log score (dLS), which approximates KL divergence, and in terms of prediction root mean square error (RMSE) --- see Section \ref{sec:matern} for details}
\label{fig:klmatern}
\end{figure}

Figure \ref{fig:maternest} shows that our scaled Fisher-scoring approach can be much more accurate than using standard Vecchia (see Section \ref{sec:matern} for details). The estimation algorithm converged quickly, requiring only around ten Fisher-scoring iterations to estimate eleven parameters.

\subsection{Prediction\label{sec:pred}}

Given the estimated parameters $\hat\bftheta$ and $\hat\bfbeta$, we would like to predict the response $y(\cdot)$ at unobserved inputs, $\bx^*_1,\ldots,\bx^*_{n_*}$.
This is equivalent to obtaining the posterior distribution of $\by^* = (y^*_1,\ldots,y^*_{n_*})^\top$, where $y^*_i = y(\bx^*_i)$. To be able to compute this distribution even for large $n$ or $n_*$, we apply a Vecchia approximation to the joint density $\dens(\by^{\text{all}})$, where $\by^{\text{all}}=(\by,\by^*)$. To do so, we employ a maximin ordering of the scaled inputs corresponding to $\by^{\text{all}}$, under the restriction that the entries of $\by$ are ordered before those in $\by^*$ \citep{Guinness2016a,Katzfuss2018}. As a result, we can write $\adens(\by,\by^*) = \adens(\by) \adens(\by^*|\by)$, where $\adens(\by)$ is as before in \eqref{eq:vecchia}, and the desired posterior predictive distribution is
\begin{equation}
    \label{eq:pred}
\textstyle\adens(\by^*|\by) = \prod_{i=1}^{n_*} \dens(y_i^*|\by^{\text{all}}_{g^*(i)}),
\end{equation}
and $g^*(i)$ contains the indices of the $m_*$ variables that are closest to $y_i^*$ (in terms of scaled distance) among those that are previously ordered in $\by^{\text{all}}$. It is possible for few or even none of the indices in a particular set $g^*(i)$ to correspond to observations, so that $y_i^*$ only conditions on other unobserved prediction variables; however, because these prediction variables may, in turn, condition on observations, the resulting predictions can be as good or better marginally (and much more accurate jointly) than predictions using only observations \citep[e.g., compare the methods LF-full and LF-ind in the first two rows of Fig.~5 in][]{Katzfuss2018}.

As in \eqref{eq:vecchia}, all the univariate conditionals, $\dens(y_i^*|\by^{\text{all}}_{g^*(i)})$, are Gaussian and can be computed in $\order(m_*^3)$ time. It is straightforward to, for example, compute the mean of or draw \emph{joint} samples from $\adens(\by^*|\by)$ using the expression \eqref{eq:pred}. In addition, $\adens(\by^*|\by)$ is jointly Gaussian with a sparse inverse Cholesky factor, from which any distributional summary of interest can be computed \citep{Katzfuss2018}.
These properties enable joint simulation and uncertainty quantification for a set of unobserved input values, such as a path through the input space.

\subsection{Variance correction\label{sec:vcf}}

Our SVecchia approach uses the GP model in Section \ref{sec:gp}, which (like virtually all statistical models) is misspecified, in that computer models are not truly realizations of such a GP. As the predictions in Section \ref{sec:pred} do not account for this model misspecification, the predictive distributions may sometimes be poorly calibrated with variances that are too small. To address this, we developed a variance-correction approach that is simple, computationally cheap, and is highly effective in the examples we have studied.

Specifically, we first estimate the parameters $\hat\bftheta$ and $\hat\bfbeta$ as described in Section \ref{sec:estimation}. We then randomly split the original training set into an ``inner'' training and test set. Computing SVecchia predictions (Section \ref{sec:pred}) at the inner test set given the inner training set, the predictive distribution for the $i$th inner test observation $y_i^\circ$ is Gaussian, say $\normal(\mu_i,\sigma^2_i)$. We modify this distribution to $\normal(\mu_i,b\sigma^2_i)$, where the variance correction factor $b$ is chosen to optimize a scoring rule \citep[e.g.,][]{Gneiting2014} involving $y_i^\circ$ and $\normal(\mu_i,b\sigma^2_i)$. In our implementation, we use the log score, which means that $b$ is chosen to minimize the sum of the negative log predictive densities at the inner test points, $-\sum_i \log \normal(y_i^\circ|\mu_i,b\sigma^2_i)$. This minimization problem consists of a simple line search, whose computational cost is negligible relative to that of estimating $\hat\bftheta$ and $\hat\bfbeta$.
Given $b$, we then make predictions at unobserved test inputs as described in Section \ref{sec:pred}, multiplying each prediction variance by $b$.

In our numerical experiments in Section \ref{sec:comparison} below, the estimates of $b$ ranged from around 1 (for the piston function) to the hundreds (for the very simplistic borehole function). The resulting corrected predictive distribution were sharp and well calibrated, with the empirical coverage of 95\% prediction intervals close to 95\%.

\subsection{Implementation\label{sec:implementation}}

We implemented our methods in \texttt{R}, building on top of the \texttt{R} package \texttt{GpGp} \citep{GpGp}. We provide the anisotropic covariance function \texttt{matern\_scaleDim} as in \eqref{eq:scaleddist}, where $\tilde K$ is the isotropic Mat\'ern covariance \citep[e.g.,][]{Stein1999}.
We also provide its special cases for half-integer smoothness values 0.5, 1.5, 2.5, 3.5, 4.5, which avoid expensive Bessel functions. Parameter estimation is based on the Fisher-scoring procedure in \texttt{GpGp}; at iterations $k=2,4,8,16,\ldots$, we update the ordering and conditioning of the current scaled inputs $\tilde\bx_1,\ldots,\tilde\bx_n$, using the exact maximin ordering algorithm implemented in \texttt{GPvecchia} \citep{GPvecchia}.
Each ordering and conditioning can be computed in quasilinear time in $n$ \citep{Schafer2020},
and in practice the added time is negligible relative to a standard Vecchia approximation that keeps the ordering and conditioning fixed.
We also provide an efficient implementation for our variance-correction procedure and for Vecchia predictions based on scaled inputs.

Due to the global nature of the Vecchia approximation (see Section \ref{sec:spatial}), it is possible to separate training of our emulator (i.e., parameter estimation) from prediction. As parameter estimation requires multiple Fisher-scoring iterations, we recommend using relatively small conditioning sets of size $m=m_{\text{est}}$ for the Vecchia density \eqref{eq:vecchia} used for the parameter estimation described in Section \ref{sec:estimation}, and of larger size $m=m_*$ for the Vecchia approximation of the predictive distribution in \eqref{eq:pred}.
In addition, our numerical experiments below showed that a random subsample of the training data of size $n_{\text{est}}$ in the low thousands was enough to estimate the small number of unknown mean and covariance parameters for the $d\leq 10$ considered here. 
For this SVecchia procedure, the computational cost is independent of the full training size $n$, aside from negligible pre-processing costs. For parameter estimation, each Fisher-scoring iteration scales roughly as $\order(n_{\text{est}}m_{\text{est}}^3)$; given the estimated parameters, prediction at $n_*$ input values based on the full training set of size $n$ scales as $\order(n_*m_*^3)$. The computations can be carried out in parallel across the $n_{\text{est}}$ terms for estimation.
Keeping in mind these computational costs, we recommend setting the tuning parameters $n_{\text{est}},m_{\text{est}},m_*$ as large as possible (to maximize accuracy) within given computational constraints. The default values in our implementation are $n_{\text{est}}=5{,}000$, $m_{\text{est}}=30$, and $m_*=140$.

The code is available at \url{https://github.com/katzfuss-group/scaledVecchia}. Using default settings, scaled-Vecchia estimation and prediction is as simple as:
\begin{verbatim}
    fit   <- fit_scaled( y.train, inputs.train )
    preds <- predictions_scaled( fit, inputs.test )
\end{verbatim}

\subsection{Design}

Our methods can also be extended straightforwardly for the design of computer experiments. 
For example, consider the following two-stage design of total size $n$. In the first stage, we obtain a small number of runs, say $n_1 = n/10$, with input values chosen by a space-filling design, such as a Latin hypercube (LH). Then, we apply our estimation method from Section \ref{sec:estimation} to the $n_1$ responses to obtain an estimate of $\bftheta$, including the estimated ranges $\hat\bflambda_1$.
In the second stage, we ``oversample'', say $N=20n$ inputs values using a LH design, and then choose the first $n_2 = n-n_1$ inputs in a maximin ordering of the scaled space determined by the range estimates $\hat\bflambda_1$ from the first stage. Finally, based on the resulting full dataset of size $n=n_1 + n_2$, we can re-estimate the parameters, and make predictions at unobserved input values as described in Section \ref{sec:pred}. Note that such a ``sensitivity-weighted distance'' has previously been considered for small sequential designs in \citet{Williams2011}.

In addition, our methods can be used for designs based on optimization criteria \citep[e.g.,][]{Mockus1989,Jones1998}, sometimes referred to as Bayesian or model-based optimization. These sequential designs at each stage require re-estimation of parameters and predictions at large numbers of inputs (e.g., to compute the expected improvement), which can be carried out rapidly using our methods.

\section{Numerical comparisons\label{sec:comparison}}

\subsection{General information\label{sec:compinfo}}

We carried out numerical studies comparing the following methods:
\begin{description}[itemsep=1pt,topsep=2pt,parsep=1pt]
\item[SVecchia:] Our proposed scaled Vecchia approximation, as described in Section \ref{sec:methodology}
\item[Vecchia:] Existing standard Vecchia approximation, with maximin ordering and nearest-neighbor conditioning based on Euclidean distance $\| \bx_i - \bx_j\|$ between inputs
\item[LowRank:] Modified predictive process \citep{Finley2009}, equivalent to Vecchia, except that all variables simply condition on the first $m$ variables in the (Euclidean) maximin ordering: $c(i) = (1,\ldots,m)$ for $i>m$
\item[laGP:] Local approximate GP \citep[][]{Gramacy2015,Gramacy2016}
\item[H-laGP:] Hybrid global-local extension of laGP \citep[][Sec.~3]{Sun2019} with pre-scaling based on a random subsample of size 1,000
\end{description}
For SVecchia, Vecchia, and LowRank, we assumed zero mean $\mu(\bx)=0$, and $K$ was assumed to be a Mat\'ern covariance with smoothness 3.5 and zero nugget. 
For each comparison, $n$ training input values were generated using Latin Hypercube sampling using the \texttt{R} package \texttt{lhs} \citep{lhs}, and $n_*$ test inputs were sampled uniformly at random on $\domain$.

\subsection{Mat\'ern simulations \label{sec:matern}}

We considered $n=5{,}000$ responses simulated from a GP with mean zero and Mat\'ern covariance function with smoothness 3.5 in $d=10$ dimensions.
We assumed two ``relevant'' input dimensions with range parameters $\lambda_1=\lambda_2=.05$, and eight less relevant inputs with range parameters $\lambda_3 = \ldots = \lambda_{10} = 5$. Only squared-exponential covariances are implemented in laGP, and so laGP was not included in this comparison. For the other three methods, we considered the average difference in log scores \citep[dLS;][]{Gneiting2014} or loglikelihoods, $\log \dens(\by) - \log \adens(\by)$, over ten datasets $\by \sim \dens(\by)$ simulated from the true model; this score approximates the KL divergence between the true and approximated model.
For each of the ten datasets and each of the approximation methods with different values of $m$, we estimated the parameters using $m_{\text{est}}=m$, made predictions at $n_*=2{,}000$ unobserved test inputs using $m_*=2m$, and computed the root mean square error (RMSE) between the true test responses $\by^*$ and the corresponding predictive means (averaged over the ten datasets).

Figure \ref{fig:klcomp} shows the dLS when assuming that the covariance function (including its parameters) was known. Vecchia was more accurate than LowRank, but SVecchia resulted in additional, substantial improvement. For example, SVecchia with $m=5$ was more accurate than Vecchia (or LowRank) with $m=50$; due to the cubic scaling in $m$, this implies a 1,000--fold decrease in computational cost for a given accuracy.
For Figure \ref{fig:maternest}, the parameters $\bftheta$ were assumed unknown and estimated from the data, but the resulting dLS were very similar to the known-parameter case.
Figure \ref{fig:maternpred} shows that SVecchia predictions were much more accurate than those using Vecchia or LowRank.

\subsection{Borehole function\label{sec:borehole}}

We carried out a simulation study comparing prediction accuracy for the Vecchia-based methods (i.e., SVecchia, Vecchia, and LowRank) using the popular borehole-function example \citep{Morris1993BayesianPrediction}, which models the water-flow rate through a borehole as a function of $d=8$ input variables.
For various training-data sizes $n$ and different values of $m$, we estimated parameters based on the training data using $m_{\text{est}}=m$, and made predictions at $n_*=2{,}000$ unobserved test inputs using $m_{\text{est}}=m$; for SVecchia, a training subsample of size $n_{\text{est}} =3{,}000$ was used for estimation if $n>n_{\text{est}}$. We computed the resulting RMSE values, averaged over ten datasets.


\begin{figure}
\centering
	\begin{subfigure}{.42\textwidth}
	\centering
	\includegraphics[width =.98\linewidth]{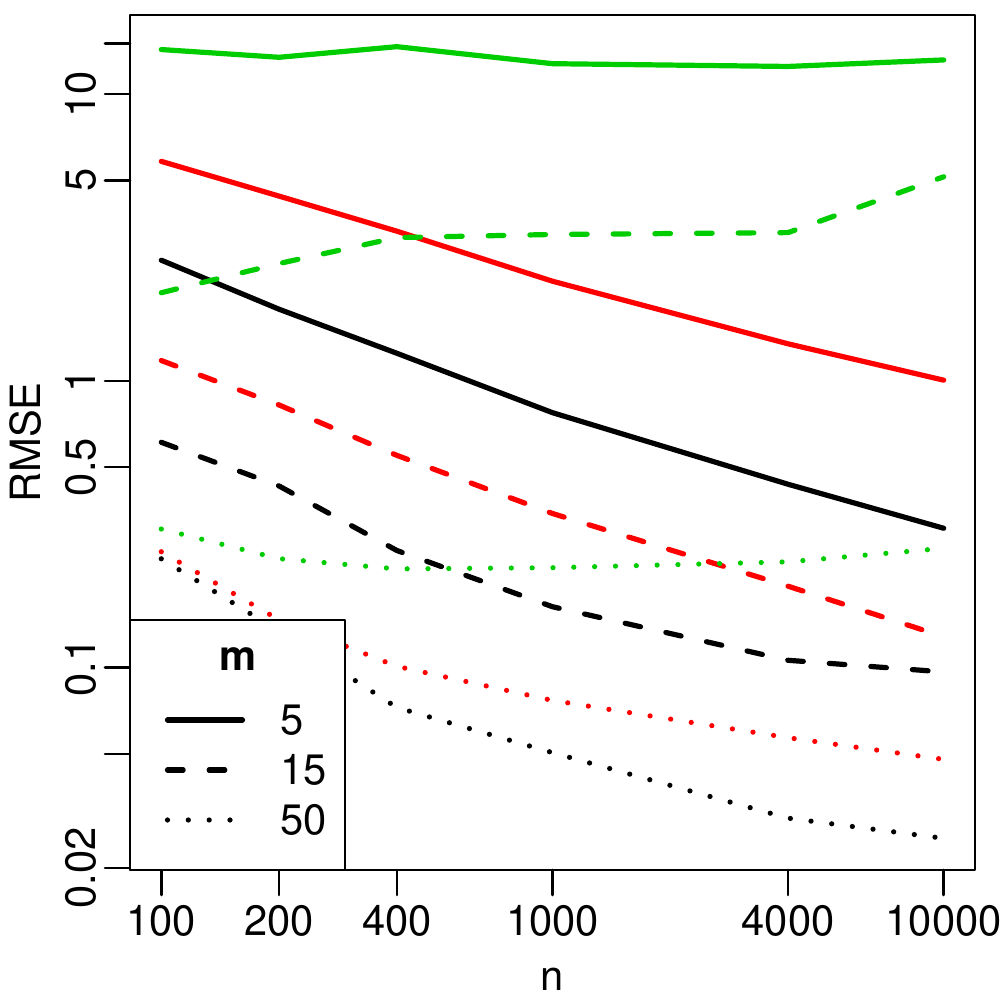}
	\caption{Increasing $n$}
	\label{fig:boren}
	\end{subfigure}%
	\begin{subfigure}{.42\textwidth}
	\centering
	\includegraphics[width =.98\linewidth]{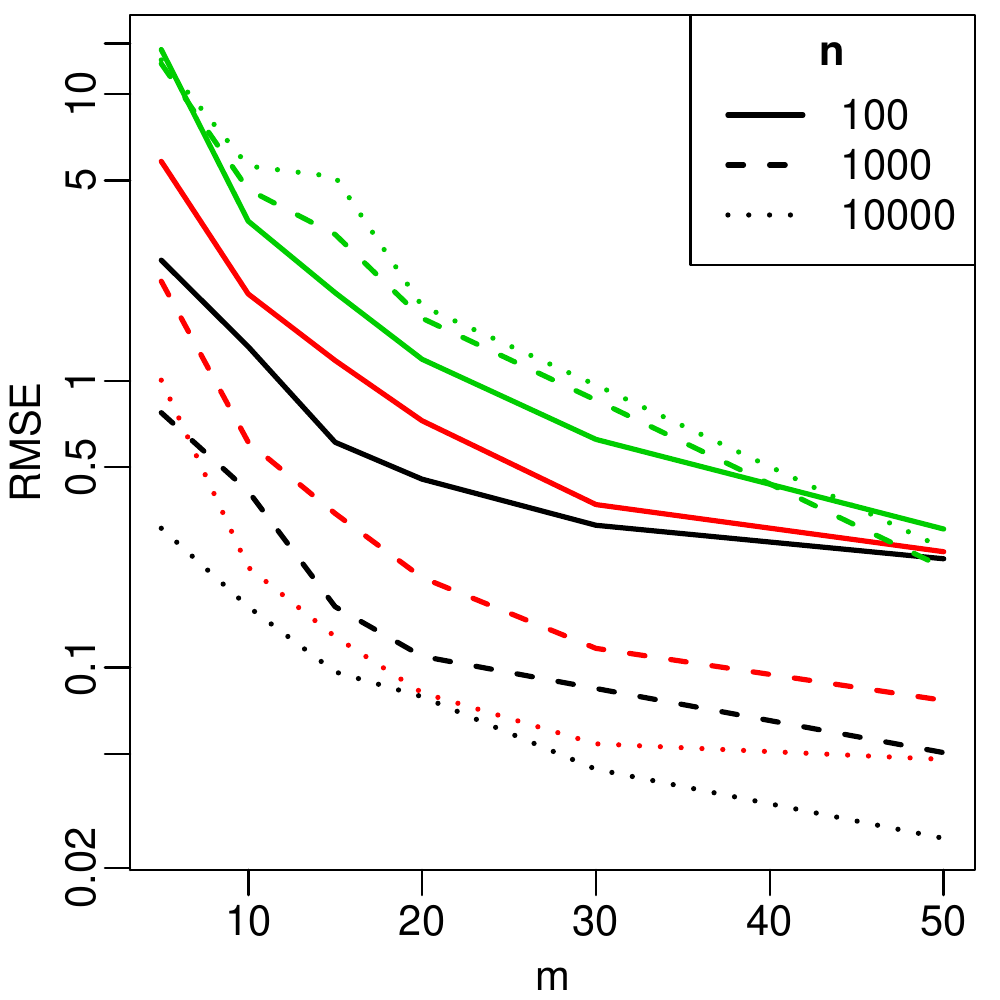}
	\caption{Increasing $m$}
	\label{fig:borem}
	\end{subfigure}%
	\begin{subfigure}{.15\textwidth}
	\centering
	\includegraphics[trim=35mm 20mm 26mm 24mm,clip,width =.98\linewidth]{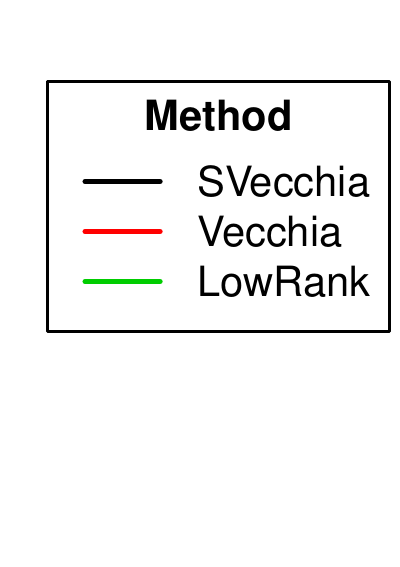} 
	\end{subfigure}%
  \caption{Root mean square error (RMSE, on a log scale) for prediction at unobserved inputs using different GP approximations for the borehole example (see Section \ref{sec:borehole} for more details)}
\label{fig:bore}
\end{figure}

The results are shown in Figure \ref{fig:bore}. For scale, the trivial predictor given by the average of the training data had an RMSE around 45, while the exact GP had an RMSE around 0.24 for $n=100$ and 0.06 for $n=400$, which was similar to the RMSE for SVecchia with $m=50$ (0.24 and 0.07). SVecchia outperformed the other approximation methods for every combination of $n$ and $m$ shown in the plots.
Note that RMSE is plotted on a log-scale. Thus, for example for $n=10{,}000$ and $m=50$, the seemingly small improvement of SVecchia over Vecchia actually corresponds roughly to a 50\% reduction in RMSE. Figure \ref{fig:boren} shows that LowRank's accuracy did not improve much with $n$, and so this method was not considered for the large-$n$ comparisons below. There is a trade-off with the tuning parameter $m$, which determines the size of the conditioning or neighbor sets: For all methods, increasing $m$ resulted in higher accuracy (Figure \ref{fig:borem}), but the computational cost also increases roughly cubically with $m$ (see Section \ref{sec:implementation} for a discussion of SVecchia's cost).

\subsection{Test functions\label{sec:testfuns}}

We then considered larger datasets generated using three physical models from the Virtual Library of Simulation Experiments \citep{Surjanovic2013}, including the borehole function from Section \ref{sec:borehole}.
We generated $n=100{,}000$ training inputs and $n_*=20{,}000$ test inputs, and averaged the results over five datasets for each test function. For (S)Vecchia, we used $m_{\text{est}} \in \{30,50\}$ and a subsample of the training data of size $n_{\text{est}} =3{,}000$ for parameter estimation, and $m_*=140$ and all training data for prediction.
For (H-)laGP, we used $30$ or $50$ neighbors for both training and prediction, and we manually specified a much smaller nugget ($10^{-7}$) than the default value to obtain more accurate predictions.
Timing results were obtained on a basic desktop computer (3.4GHz Intel Quad Core i5-3570), using one core (single-threaded) for (S)Vecchia and using all four cores for (H-)laGP. 

\begin{figure}
\centering
~ \hfill 
\includegraphics[trim=0mm 5mm 0mm 5mm,clip,width =.5\linewidth]{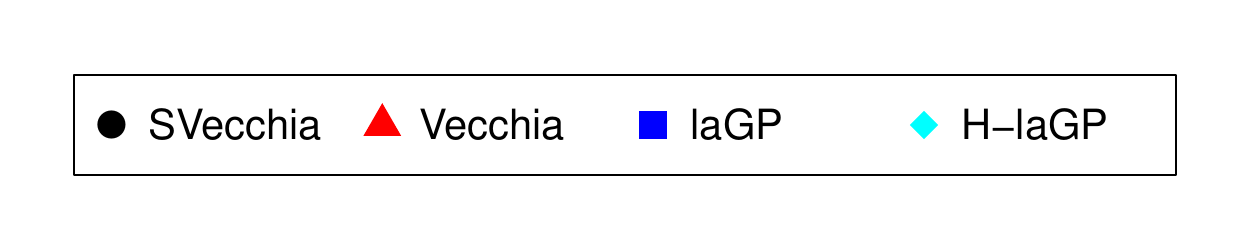} 
\hfill
\includegraphics[trim=0mm 5mm 0mm 5mm,clip,width =.2\linewidth]{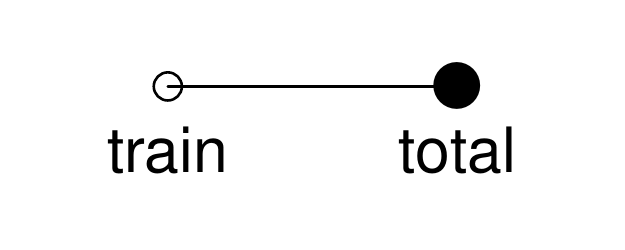}
\hfill ~
\\
	\begin{subfigure}{.33\textwidth}
	\centering
	\includegraphics[width =.98\linewidth]{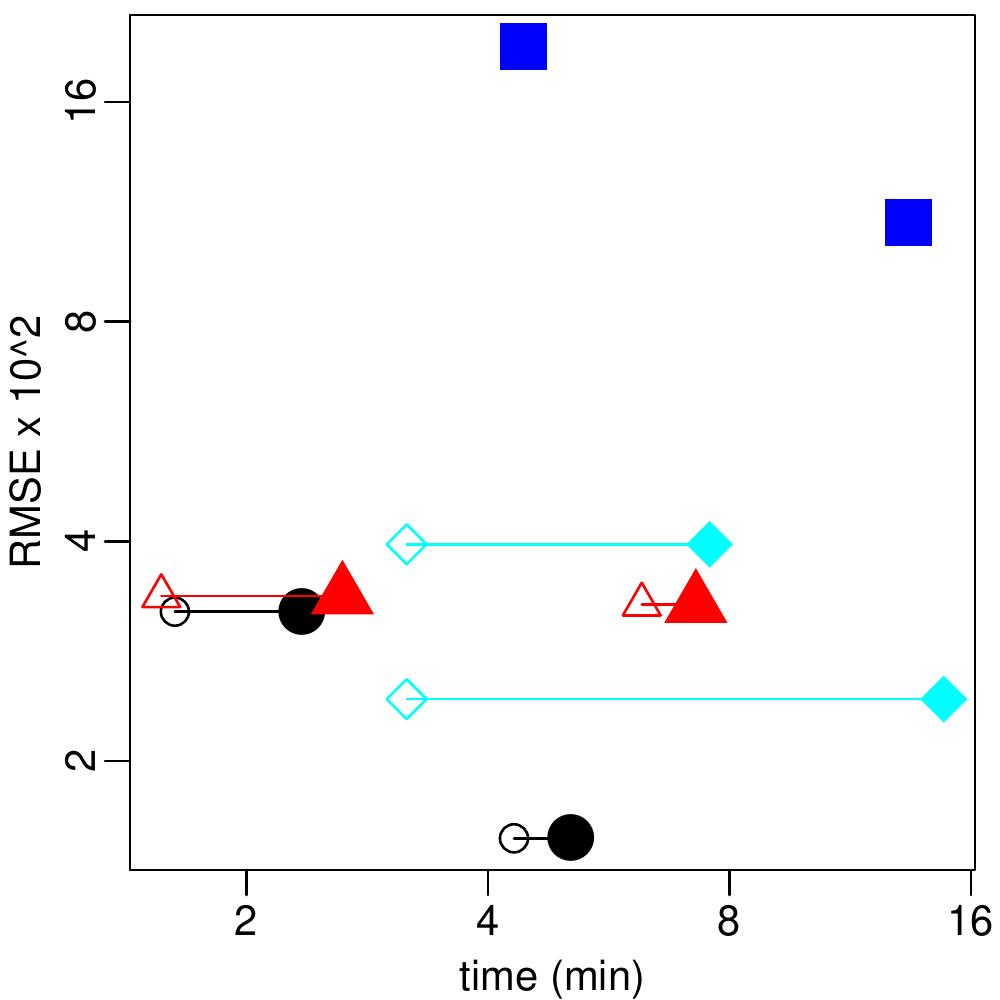}
	\caption{borehole ($d=8$)}
	\label{fig:bore100}
	\end{subfigure}%
	\begin{subfigure}{.33\textwidth}
	\centering
	\includegraphics[width =.98\linewidth]{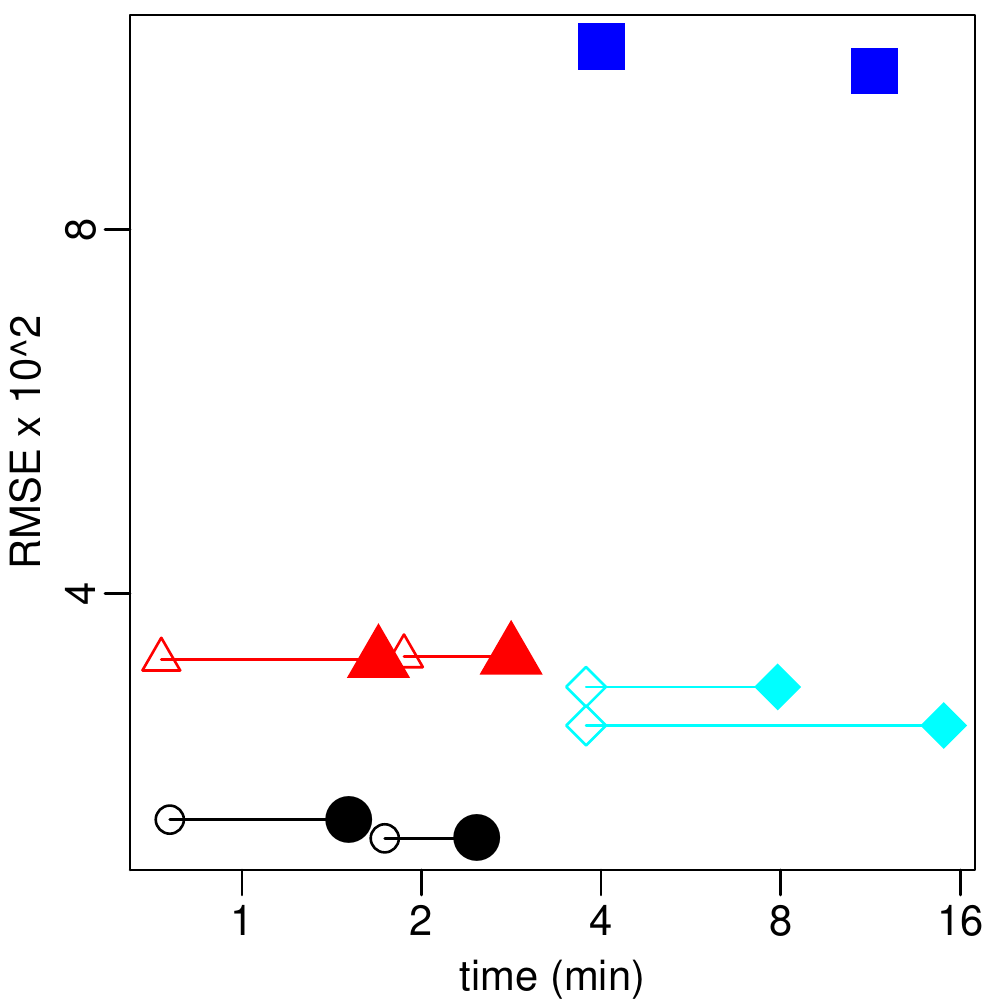}
	\caption{robot arm ($d=8$)}
	\label{fig:robot}
	\end{subfigure}%
	\begin{subfigure}{.33\textwidth}
	\centering
	\includegraphics[width =.98\linewidth]{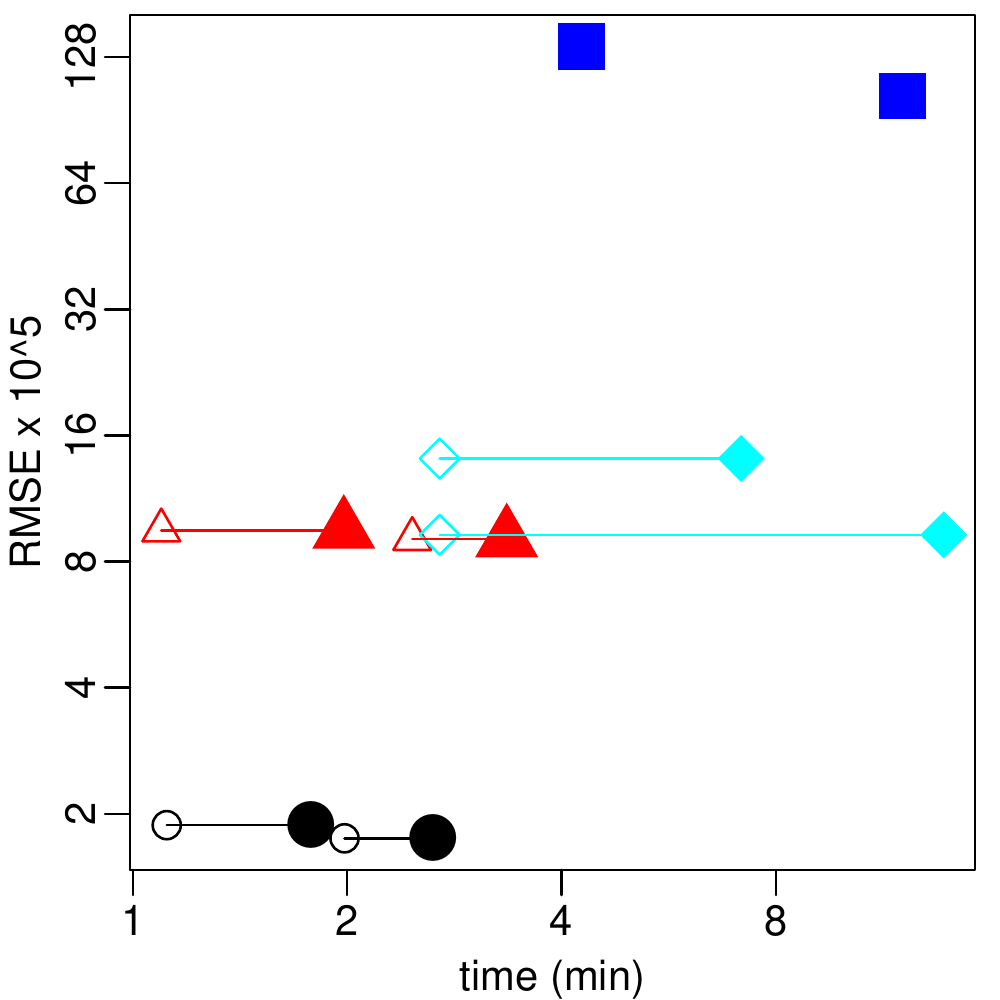}
	\caption{piston ($d=7$)}
	\label{fig:piston}
	\end{subfigure}%
  \caption{Comparison of root mean square error (RMSE) versus computing time (both on a log scale) for test functions in Section \ref{sec:testfuns}, with two different tuning-parameter settings for each function and method. We provide training and total (i.e., training plus prediction) times, on a single core for (S)Vecchia and on four cores for (H-)laGP.}
\label{fig:testfuns}
\end{figure}

The results are summarized in Figure \ref{fig:testfuns}, and the detailed numerical results are given in Table \ref{tab:testfuns} in Appendix \ref{app:testfuns}. For all three test functions, SVecchia was the most accurate, despite having the lowest computational cost. The cost of recomputing the ordering and conditioning sets for SVecchia at certain Fisher-scoring iterations was negligible, as it only took a fraction of a second.

To assess the accuracy of the uncertainty quantification, we also computed several scores for the predictive distributions and the implied 95\% prediction intervals for the piston test function, using the $m_{\text{est}}=30$ and $30$-neighbor setting from above. The interval score, log score, and continuous ranked probability score (CRPS) each simultaneously quantify calibration and sharpness of the marginal predictive distributions \citep[see, e.g.,][for details]{Gneiting2014}. The energy score \citep{Gneiting2008a} assesses the quality of the joint predictive distribution at all $n_*=20{,}000$ test inputs, which were only available for (S)Vecchia. The (S)Vecchia intervals were well calibrated. SVecchia performed by far the best in terms of all other scores.

\begin{table}
\centering \footnotesize
\begin{tabular}{l|rrrrrr}
 & ICov (\%) & IWidth & IScore & LogScore & CRPS & Energy \\ 
  \hline
SVecchia & 95.4 & 2.9 & 9.8 & -97.8 & 0.7 & 182.0 \\ 
  Vecchia & 95.4 & 14.7 & 51.5 & -81.8 & 3.7 & 977.0 \\ 
  laGP & 81.3 & 81.3 & 742.8 & -57.7 & 47.5 &  \\ 
  H-laGP & 98.5 & 31.0 & 70.0 & -76.7 & 6.6 & 
\end{tabular}
\caption{Scores evaluating the accuracy of the uncertainty quantification of the predictive distributions for the piston test function. Lower is better for all scores except ICov. All scores except ICov and LogScore were multiplied by $10^5$. ICov = empirical coverage of 95\% prediction intervals; IWidth = average interval width; IScore = interval score; CRPS = continuous ranked probability score; Energy = energy score. }
\label{tab:testUQ}
\end{table}

In general, it is difficult to set a comparison in which all methods are placed on perfectly equal footing. The Vecchia approaches used a subsample for parameter estimation, and the full training set with a larger conditioning-set size $m_*$ for prediction. H-laGP also uses data for global pre-estimation, but in a different manner. Both laGP methods do some estimation on the fly. For all methods, increasing the size of the conditioning or neighbor sets improves the accuracy but also increases the computational cost. Timing results will also depend heavily on a number of other factors, including $n_{\text{est}}$, $n_*$, implementation, parallelization, and the computing environment. Due to the good parallelization properties of the laGP implementation, laGP prediction times could potentially be pushed below those of single-core SVecchia by using enough cores for laGP.

While the test functions are smooth, deterministic functions without noise, we also tested estimating a noise variance using SVecchia. For the piston function, we artificially added observation noise with variance $\tau^2$, with $\tau = .02$. When including the noise variance as an unknown parameter to estimate in the Fisher scoring algorithm, we obtained a highly accurate estimate of $\hat\tau = .0198$, even with a small $m_{\text{est}}=30$.

\subsection{Computer model for satellite drag\label{sec:satellite}}

Finally, we carried out comparisons using a computer simulator for atmospheric drag coefficients of satellites in low Earth orbit under varying input conditions. A detailed description of the computer model and a previous analysis using state-of-the-art GP emulators can be found in \citet{Sun2019}, with data and results available at \url{https://bitbucket.org/gramacylab/tpm/src/master/}. In short, we considered simulations of drag coefficients for the Hubble space telescope with $d=8$ inputs. The simulation runs consist of $n=2\times 10^6$ responses for each of six pure chemical species, which can be combined into actual drag coefficients by computing a weighted average of the species.

\begin{figure}
\centering
	\begin{subfigure}{.45\textwidth}
	\centering
	\includegraphics[width =.92\linewidth]{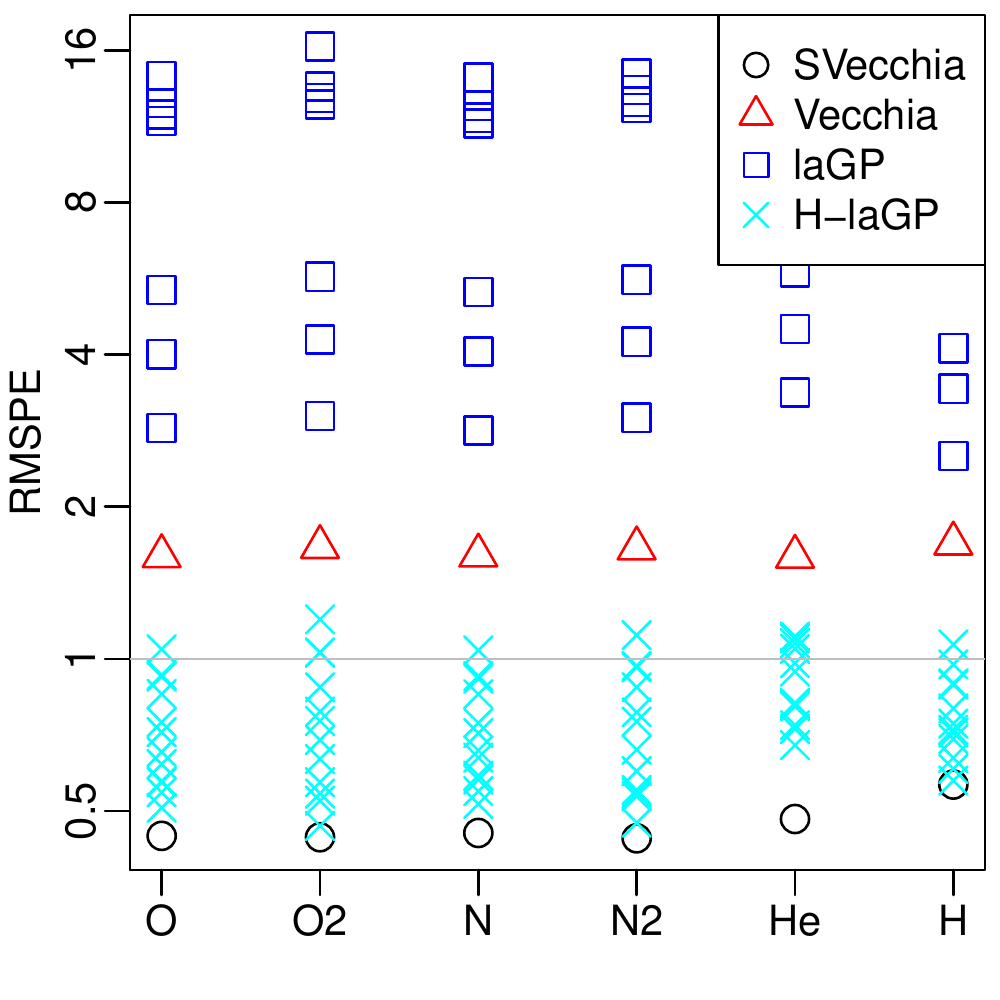}
	\caption{CV root mean square percentage error (RMSPE)}
	\label{fig:satCV}
	\end{subfigure}%
	\begin{subfigure}{.45\textwidth}
	\centering
	\includegraphics[width =.92\linewidth]{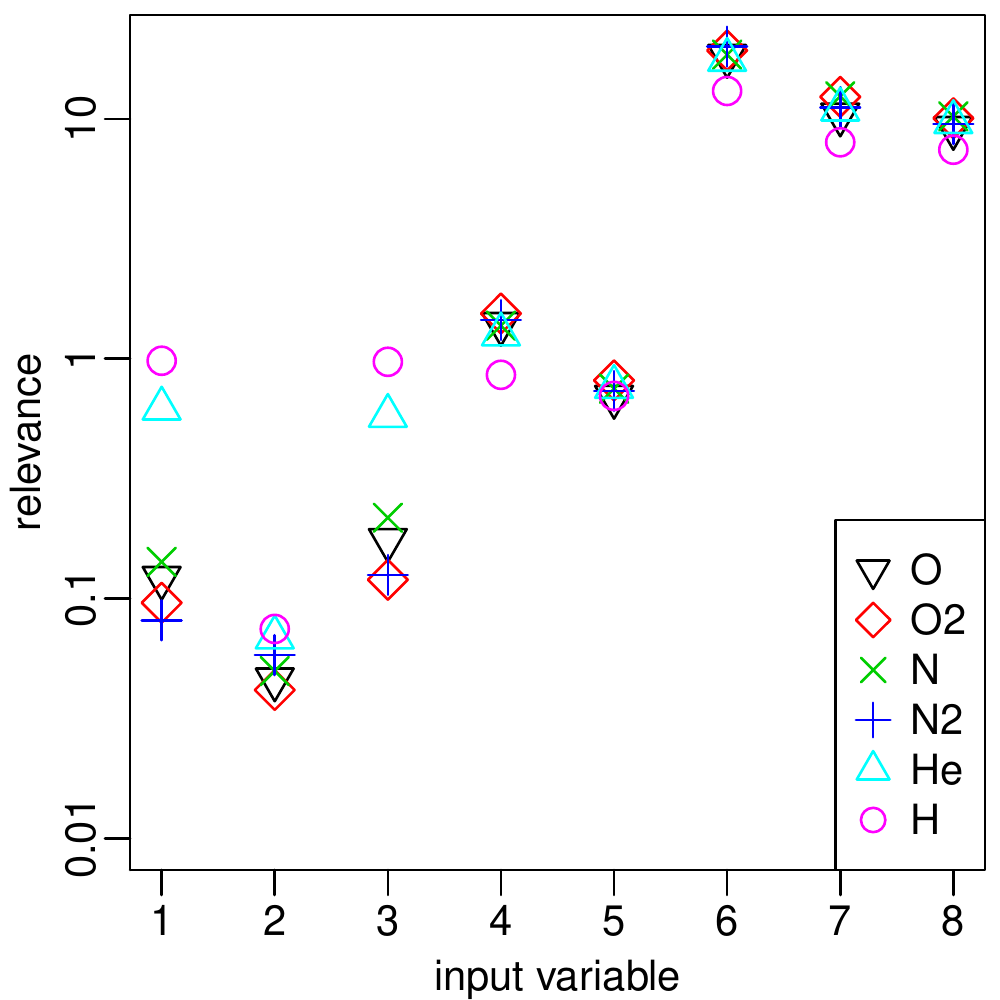}
	\caption{SVecchia estimates of relevance of input variables}
	\label{fig:satrelevance}
	\end{subfigure}%
  \caption{Results for the six chemical species in the satellite-drag simulator. H-laGP: hybrid global-local extensions of laGP. Relevance: $1/\hat\lambda_l$ (see details below \eqref{eq:scaledinputs})}
\label{fig:sat}
\end{figure}

As in \citet[][Sect.~6.1]{Sun2019}, we carried out 10-fold cross-validation (CV), separately for each of the six species. For the Vecchia-based methods, we used $m_*=140$ for prediction, and we used $m_{\text{est}}=30$ and a randomly selected subset of size $n_{\text{est}} =10{,}000$ for parameter estimation. We also tried estimation using the full dataset (i.e., $n_{\text{est}}=n$) and a larger $m_{\text{est}}$, but the increase in predictive accuracy was small relative to the increase in computational cost. The parameter estimates were quite stable between different CV folds. One example of the estimated relevance $1/\hat\lambda_l$ is shown in Figure \ref{fig:satrelevance} for each input variable $x_l$ and each species; the highest and lowest relevance differed by two-to-three orders of magnitude, indicating that SVecchia's corresponding scaling of the input dimensions should be useful for emulating this simulator. 

Figure \ref{fig:satCV} shows a comparison of CV prediction accuracy in terms of root mean square percentage error (RMSPE). We compared the Vecchia-based methods to the 19 laGP variants considered and described in \citet{Sun2019}, seven of which are versions of the basic, local-only laGP, and twelve of which are hybrid global-local laGP (H-laGP) extensions. Vecchia was more accurate than the basic laGP methods, but none of these approaches was able to achieve the standard benchmark of a 1\% relative error, indicated by the horizontal line. 
In contrast, SVecchia met the benchmark and was the most accurate method for all six chemical species. While the accuracy improvement might look small on the log scale of Figure \ref{fig:satCV}, note that the RMSPE of the best-performing H-laGP method (``alcsep2.sb'') in Figure \ref{fig:satCV} was considerably higher than the SVecchia RMSPE for several species, 
ranging from roughly 2\% higher for H, to around 14\% for O and N, up to 40\% for He.
This is especially remarkable when considering that the total time for estimation and prediction for SVecchia was only around 13 to 14 minutes (4--5min for estimation and roughly 9min for prediction) per species and fold, on a single core on a basic desktop computer; the best-performing laGP method took up to 45 core hours according to \citet{Sun2019}, which is around 200 times as long.

We also examined predictions along likely trajectories in low Earth orbit, which corresponds to paths in input space. \citet[][Sect.~6.2]{Sun2019} consider two trajectories, for a quiet and active regime, each for $n_* = 8{,}600$ ten-second intervals (i.e., about one day). Predictions are made for each of the six pure chemical species, which are then averaged according to weights corresponding to the actual chemical compositions for each of the two regimes. Given estimated parameters, joint prediction using SVecchia scales linearly in $n_*$, the number of test inputs. Thus, SVecchia can produce joint predictions (e.g., samples from the joint predictive distribution) for the day-long trajectory with $n_*=8{,}641$ in less than one minute on a single core; this is less time than it takes the most accurate laGP method (``ALC-ex'') to compute predictions for small subsets of size $n_* = 100$.
The RMSPE for ALC-ex was about 39\% and 8\% higher than for SVecchia for the quiet and active regimes, respectively. However, the trajectories traverse only a small fraction of the input space, so that comparing prediction scores for only two such trajectories is not statistically meaningful. 
Vecchia even happened to have a smaller RMSPE than SVecchia for the active regime.

\begin{table}
\centering \footnotesize
\begin{tabular}{l|rrrrrrr}
 & RMSPE & ICov & IWidth & IScore & LogScore & CRPS & Energy \\ 
  \hline
SVecchia & 38.3 & 95.8 & 1.9 & 5.2 & -32.9 & 0.5 & 6.7 \\ 
  Vecchia & 137.7 & 96.8 & 9.5 & 25.9 & -18.5 & 2.0 & 28.7
\end{tabular}
\caption{For the O species in the satellite-drag application, scores (same as in Table \ref{tab:testUQ}) for joint predictions at 100 pseudo-trajectories of size 100 each. LogScore was multiplied by 10, and all other scores were multiplied by 100. The scores for the other five chemical species were very similar.}
\label{tab:pseudo}
\end{table}

For a more statistically meaningful comparison, we created 100 pseudo-trajectories, each of which was of size 100 and obtained by randomly selecting one of the $n$ inputs and then sequentially selecting the nearest input. We computed joint predictions from (S)Vecchia at each pseudo-trajectory, using the remaining $n-100^2$ observations as training data. The resulting scores, averaged over the 100 pseudo-trajectories, are shown in Table \ref{tab:pseudo}; SVecchia strongly outperformed Vecchia.
We did not have predictions for the (H-)laGP methods from Figure \ref{fig:satCV} for this experiment; we instead considered a comparison to the (H-)laGP methods from Section \ref{sec:testfuns}, but these methods were not tuned to this satellite-drag application and their scores were not competitive.

\section{Conclusions and future work\label{sec:conclusions}}


We have introduced a fast and accurate scaled-Vecchia approximation for Gaussian-process emulation of large computer experiments. The Vecchia approach relies on an ordered conditional approximation, which results in a joint global likelihood and natural joint prediction and uncertainty quantification. Maximin ordering ensures that high accuracy can be achieved by simply conditioning on (previously ordered) nearest neighbors. For the high input dimensions prevalent in computer experiments, our approach applies the Vecchia approximation in a scaled input space, for which the scaling parameters are automatically determined from the data using a fast parameter-estimation procedure. For fixed conditioning-set sizes, this estimation procedure requires linear time in the number of estimation data, while joint prediction scales linearly in the number of prediction points.

In several numerical comparisons, our proposed method substantially outperformed existing approximations, in that it was able to produce more accurate results in less computational time. For example, for the satellite-drag computer simulator, even a basic version of scaled Vecchia was more accurate and several orders of magnitude faster than the state-of-the-art laGP approaches.
As it can produce highly accurate joint predictions with a few lines of code in minutes on modest computers even for big datasets, we consider scaled Vecchia to be a good candidate for a default approach for emulating large computer experiments.


Additional improvements in prediction accuracy may be possible for our method by considering nonstationary covariance functions, such as a Mat\'ern covariance whose parameters vary over input space \citep{Paciorek2006}; ordering and conditioning should then be correlation-based (Kang \& Katzfuss, in prep.). Such a correlation-based approach would also be possible for joint emulation for multivariate or functional computer-model output.

More sophisticated frequentist variable (i.e., input-dimension) selection could be achieved by adding a lasso-type L1 penalty for the inverse range parameters to \eqref{eq:fisherdens}. MCMC-based Bayesian inference can also be accurately approximated using Vecchia approaches \citep[][App.~E]{Finley2017,Katzfuss2017a}; straightforward extensions include scaling the input space at certain MCMC iterations, and variable selection \citep{linkletter2006variable}.

Non-Gaussian computer-model responses could be analyzed by combining scaled Vecchia with the Vecchia-Laplace approximation of generalized GPs \citep{Zilber2019}.
Finally, it would be interesting to investigate the use and extension of our methods in the context of computer-model calibration \citep{Kennedy2001}.

\footnotesize
\appendix
\section*{Acknowledgments}
Katzfuss's research was partially supported by National Science Foundation (NSF) Grants DMS--1654083, DMS--1953005, CCF--1934904, and by a Texas A\&M University System National Laboratories Office grant on ``Scalable Gaussian-Process Methods for the Analysis of Computer Experiments.''
Lawrence's research was supported by the Laboratory Directed Research and Development program of Los Alamos National Laboratory under project number 20200065DR. Guinness's research was supported by the NSF under grant No.\ 1916208 and the National Institutes of Health under grant No.\ R01ES027892. We would like to thank Bobby Gramacy and Furong Sun for providing help, data, and laGP results for the application in Section \ref{sec:satellite}.

\section{Detailed test-function comparison results\label{app:testfuns}}

Table \ref{tab:testfuns} provides the specific accuracy and timing results underlying Figure \ref{fig:testfuns}, as discussed in Section \ref{sec:testfuns}.

\begin{table}[htbp]
\centering \footnotesize
\begin{tabular}{l|r|r|r|r|r|r}
 &   \multicolumn{2}{c|}{borehole ($d=8$)}   & \multicolumn{2}{c|}{robot arm ($d=8$)} & \multicolumn{2}{c}{piston ($d=7$)} \\  
method & E$\times 10^2$ & time (min) & E$\times 10^2$ & time (min) & E$\times 10^5$ & time (min) \\ 
  \hline
SVecchia (30/140) & \textcolor{red}{3.2} & 1.6$+$0.7$=$ \textcolor{red}{2.3} & \textcolor{red}{2.6} & 0.8$+$0.8$=$ \textcolor{red}{1.6} & \textcolor{red}{1.9} & 1.1$+$0.7$=$ \textcolor{red}{1.8} \\ 
  Vecchia (30/140) & 3.4 & 1.6$+$1.1$=$ 2.7 & 3.5 & 0.7$+$1.0$=$ 1.7 & 9.5 & 1.1$+$0.9$=$ 2.0 \\ 
  laGP (30) & 19.0 & 0$+$4.4$=$ 4.4 & 11.3 & 0$+$4.0$=$ 4.0 & 135.4 & 0$+$4.3$=$ 4.3 \\ 
  H-laGP (30) & 4.0 & 3.2$+$4.3$=$ 7.5 & 3.3 & 3.9$+$4.0$=$ 7.9 & 14.1 & 3.1$+$4$=$ 7.1 \\ 
  \hline
  SVecchia (50/140) & \textcolor{red}{1.6} & 4.3$+$0.8$=$ \textcolor{red}{5.1} & \textcolor{red}{2.5} & 1.7$+$0.7$=$ \textcolor{red}{2.4} & \textcolor{red}{1.7} & 2$+$0.7$=$ \textcolor{red}{2.7} \\ 
  Vecchia (50/140) & 3.3 & 6.2$+$1$=$ 7.2 & 3.5 & 1.9$+$1.0$=$ 2.9 & 9.1 & 2.5$+$0.9$=$ 3.4 \\ 
  laGP (50) & 10.9 & 0$+$13.4$=$ 13.4 & 10.8 & 0$+$11.5$=$ 11.5 & 103.2 & 0$+$12.0$=$ 12.0 \\ 
  H-laGP (50) & 2.4 & 3.1$+$11.7$=$ 14.8 & 3.1 & 3.7$+$11.3$=$ 15.0 & 9.3 & 2.3$+$11.5$=$ 13.8
\end{tabular}
\caption{Comparison for test functions in Section \ref{sec:testfuns}. E: root mean square error. Computing times for training $+$ prediction $=$ total, on one core for (S)Vecchia and on four cores for (H-)laGP. Numbers after method names are $(m_{\text{est}}/m_*)$ for (S)Vecchia, and $(\text{neighborhood size})$ for (H-)laGP. Smallest errors and computing times are highlighted in red for each test function and for the first and last four rows, respectively.}
\label{tab:testfuns}
\end{table}

\bibliographystyle{apalike}
\bibliography{mendeley,additionalrefs}

\end{document}